\documentclass[aps,prl,twocolumn,floatfix,showpacs]{revtex4-1}
\usepackage{epsf,graphicx,amsmath,amssymb,verbatim}
\newcommand{\be}{\begin{equation}}
\newcommand{\ee}{\end{equation}}

\begin{document}
\title{Correlated couplings and robustness of coupled networks}
\author{Won-kuk~Cho}
\affiliation{Department of Physics, Korea University, Seoul 136-713, Korea}
\author{K.-I.~Goh}
\thanks{Email: kgoh@korea.ac.kr}
\affiliation{Department of Physics, Korea University, Seoul 136-713, Korea}
\author{I.-M.~Kim}
\affiliation{Department of Physics, Korea University, Seoul 136-713, Korea}
\date{\today}
\begin{abstract}
Most real-world complex systems can be modelled by coupled networks
with multiple layers.
How and to what extent the pattern of couplings between network layers
may influence the interlaced structure and function of coupled networks
are not clearly understood.
Here we study the impact of correlated inter-layer couplings on
the network robustness of coupled networks using percolation concept.
We found that the positive correlated inter-layer coupling enhances
network robustness in the sense that it lowers the percolation
threshold of the interlaced network than the negative correlated coupling case.
At the same time, however,
positive inter-layer correlation leads to smaller giant component
size in the well-connected region, suggesting potential disadvantage 
for network connectivity, as demonstrated also with
some real-world coupled network strucutres.

\end{abstract}
\maketitle

In the past decade, network has proved to be a useful framework
to model structural complexity of complex systems \cite{small,scale}.
By abstracting a complex system into nodes (constituents)
and links (interactions between them), the resulting graph 
could be efficiently treated analytically and numerically,
through which a large body of new physics of complex systems 
has been acquired \cite{caldarelli-book,havlin-book}.
Most studies until recently have focused on the properties
of isolated, single networks, such as the World wide web, 
coauthorship network of a specific subject, and the metabolic 
network of a microorganism.
In most, if not all, situations, however, a network in question
is merely a part of a larger system; world wide web could only
function properly in conjuction with physical network of routers,
a researcher could take part in multiple social networks 
via various professional or informal relationships other than
coauthorship relations, and a protein in a cell could function
both as a metabolic enzyme and by binding with other proteins,
thereby taking part in both metabolic and protein networks.
Therefore, a more complete description of complex systems
would be a system of networks coupled one another.
Typical forms of network coupling would be layered structure \cite{kurant},
multiplexity of links \cite{multiplexity}, and interaction \cite{leicht}
or interdependency \cite{buldyrev} between network layers.

Only recently, such coupled network structures have started to be
studied actively \cite{leicht,buldyrev,more,buldyrev2}. 
Leicht and D'Souza \cite{leicht} studied what they called 
interacting networks, in which two networks are coupled via inter-network edges,
and developed a generating function formalism to study their percolation 
properties.
Buldyrev {\it et al.}~\cite{buldyrev} studied the interdependent networks, 
in which mutual connectivity in two network layers plays an important role, 
and found that catastrophic cascades of failure generically occur 
due to the interdependency.
In both studies, network layers are coupled in an uncorrelated way, 
in the sense that
the connections or pairings between nodes in different layers are 
taken to be random. In real complex systems, however, 
correlated inter-layer couplings can be siginificant.
For example, one may expect that a person with many friends in one
social network would also have many friends in another social network,
being a friendly person.
Such correlated coupling may alter the interlaced connectivity structure 
and large-scale properties of dynamic processes on the coupled networks, 
such as network robustness \cite{cohen}
or cascading failures \cite{buldyrev,buldyrev2,goh}.
Therefore, the impact of correlated inter-layer coupling to system's structure 
and function needs to be understood.

In this paper, we study how the network robustness is affected
by the correlated inter-layer coupling in two-layer coupled networks.
Specifically, we consider two networks with identical set of $N$ nodes.
They can be individuals participating in two different social networks
or proteins with activities via enzymes and multi-protein assembly.
Each layer $l$ is specificed with the intra-layer
degree distribution $\pi^{(l)}(k_l)$, where degree $k_l$ is the
number of links within layer $l$ (intra-layer links) of the node.
The coupled network can be specified by the joint degree distribution
$\Pi(k_1,k_2)$. Generalization into $l>2$ layers is straightforward,
and in this study we restrict ourselves to $l=2$ case.
Next the network layers are overlaid one another, and the interlaced
network is constructed. The degree of a node in the interlaced network
is given by $k=k_1+k_2-k_{o}$, where $k_o$ denotes the number of
links overlapping in the two layers, which can be neglected
in sparse networks. Therefore, the generating function $G_0(x)$ for 
the degree distribution $P(k)$ of the interlaced network can be written as
\be
G_0(x)=\sum_{k=0}^{\infty}P(k)x^{k}=\sum_{k_1,k_2}\Pi(k_1,k_2)x^{k_1+k_2}~.
\ee
The primary quantity of interest is the size of giant component $S$
of the interlaced network, given by the fraction of nodes belonging
to the largest component in the limit $N\to\infty$.
Following standard generating function technique \cite{newman},
$S$ can be obtained via 
\be S=1-G_0(x^*), \ee
with $x^*$ being the solution of $x=G_1(x)$, 
where $G_1(x)=\frac{1}{\langle k\rangle}\frac{d}{dx}G_0(x)$
is the generating function of the probability distribution
of remaing degrees $q=k-1~(k\ge1)$.
The condition for existence of the giant component (that is, $S>0$)
is given by the Molloy-Reed criterion \cite{molloy}, 
$\sum_kk(k-2)P(k)>0$. Therefore, understanding of
how the inter-layer coupling correlation encoded in the joint degree distribution 
$\Pi(k_1,k_2)$ translates into the degree distribution $P(k)$ of
the interlaced network can provide critical insight on network robustness.

In the absence of inter-layer correlations, the joint degree
distribution factorizes, 
$\Pi_{uncorr}(k_1,k_2)=\pi^{(1)}(k_1)\pi^{(2)}(k_2)$,
and therefore the degree distribution of the interlaced network 
is given by the convolution of $\pi^{(l)}(k_l)$,
$P_{uncorr}(k)=\sum_{k_1}\pi^{(1)}(k_1)\pi^{(2)}(k-k_1)$,
and its generating function $G_0^{uncorr}(x)=g_0^{(1)}(x)g_0^{(2)}(x)$,
where $g_0^{(l)}(x)$ is the generating function of $\pi^{(l)}(k_l)$.
In the presence of inter-layer correlations, however,
$P(k)$ is related with $\pi^{(l)}(k_l)$ in a nontrivial way. 

\begin{figure}[t]
\centerline{\epsfxsize=.70\linewidth \epsfbox{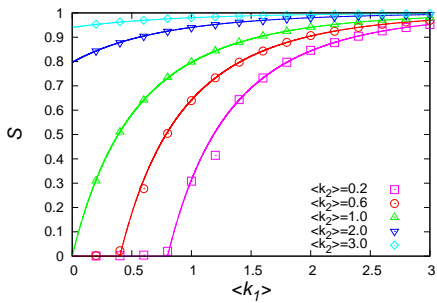}}
\caption{(Color Online) The giant component size $S$
of interlaced networks of two ER networks in the absence of inter-layer
correlations is plotted 
as a function of $\langle k_1\rangle$ for various $\langle k_2\rangle$.
Nmerical simulation results obtained with $N=10^4$ nodes (points)
and theoretical curves obtained by the generating function
method (solid lines) are shown together.
}
\end{figure}

To gain more insight, we consider specific 
network models. First we consider two layers of Erd\H{o}s-R\'enyi (ER)
networks with mean degrees $\langle k_1\rangle$ and $\langle k_2\rangle$, 
respectively. For the uncorrelated case, one can proceed with
exact calculations as 
$G_0(x)=e^{(\langle k_1\rangle+\langle k_2\rangle)(x-1)}$,
from which $S$ can be obtained using Eq.~(2). 
The interlaced network is nothing but another ER network with mean degree
$\langle k\rangle=\langle k_1\rangle+\langle k_2\rangle$.
Numerical simulations with two-layer ER networks with $N=10^4$
perfectly confirmed this theoretical prediction (Fig.~1).
Percolation transition occurs when $\langle k_1\rangle+\langle k_2\rangle=1$.

Next we introduce the correlated inter-layer coupling in the two-layer networks.
For simplicity, we consider two extreme cases of maximally-positive and maximally-negative 
correlations. In the maximally-positive inter-layer correlation, 
the ranks in degrees of a node in the two layers are equal, meaning
that the highest degree node in one layer is also the highest degree
node in the other layer, and similarly for the lowest degree nodes.
Specifically, the Spearman rank correlation coefficient $r_s$ of the degrees of
a node, defined by 
\be
r_s=1-\frac{6\sum_{i=1}^N\Delta_i^2}{N(N^2-1)}~,
\ee
where $\Delta_i$ is the difference in the ranks of node $i$'s
degrees in the two layers, is $r_s^{(pos)}\approx1$. 
In the maximally-negative case, it is the opposite that
the highest degree node in one layer has the lowest degree in the
other layer, having the reverse ordering of ranks. In this case,
the rank correlation coefficient is $r_s^{(neg)}\approx-1$
\cite{foot}.

To construct the networks with maximal correlations,
we arrange the nodes in the order of their degrees in each layer
and match two nodes from each layer in order of the degree-rank
(maximally-positive) or in the opposite order of the degree-rank
(maximally-negative).
We performed numerical simulations following this procedure
for the two-layer ER networks and measured the size of giant
component as a function of $\langle k_1\rangle$ with fixed
$\langle k_2\rangle$ (Fig.~2, points). We also performed the following
mean-field-like numerical calculations for $P(k)$ for 
both cases and used them to obtain $S$ through Eqs.~(1-2), 
which are found to agree with the numerical simulation results
remarkably well (Fig.~2, lines).
For maximally-positive case, we setup two ordered sequences 
obtained from the cumulative intra-layer
degree distribution, $\pi_{cumul}^{(l)}(k_l)=\sum_{k=0}^{k_l}\pi^{(l)}(k)$,
of each layer $l$,
and merge them to make another ordered sequence. 
This sequence is nothing but the cumulative distribution of $P(k)$,
from which we can calculate $P(k)$ easily.
For maximally-negative case, we need to use two ordered sequences
given by the cumulative degree distribution in one layer
and the complementary cumulative degree distribution,
$\pi_{comple}^{(l)}(k_l)=1-\sum_{k=0}^{k_l}\pi^{(l)}(k)$, in the other layer.

\begin{figure}
\centerline{\epsfxsize=.70\linewidth \epsfbox{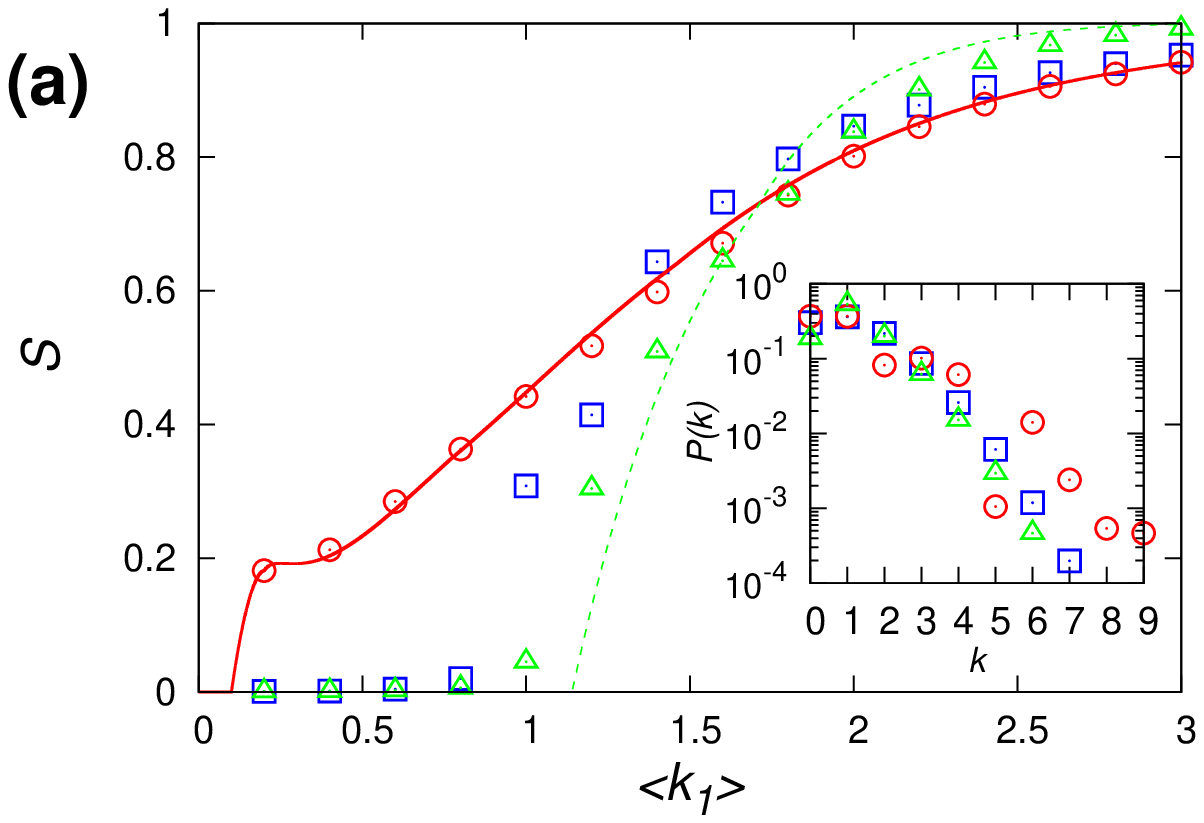}}
\centerline{\epsfxsize=.70\linewidth \epsfbox{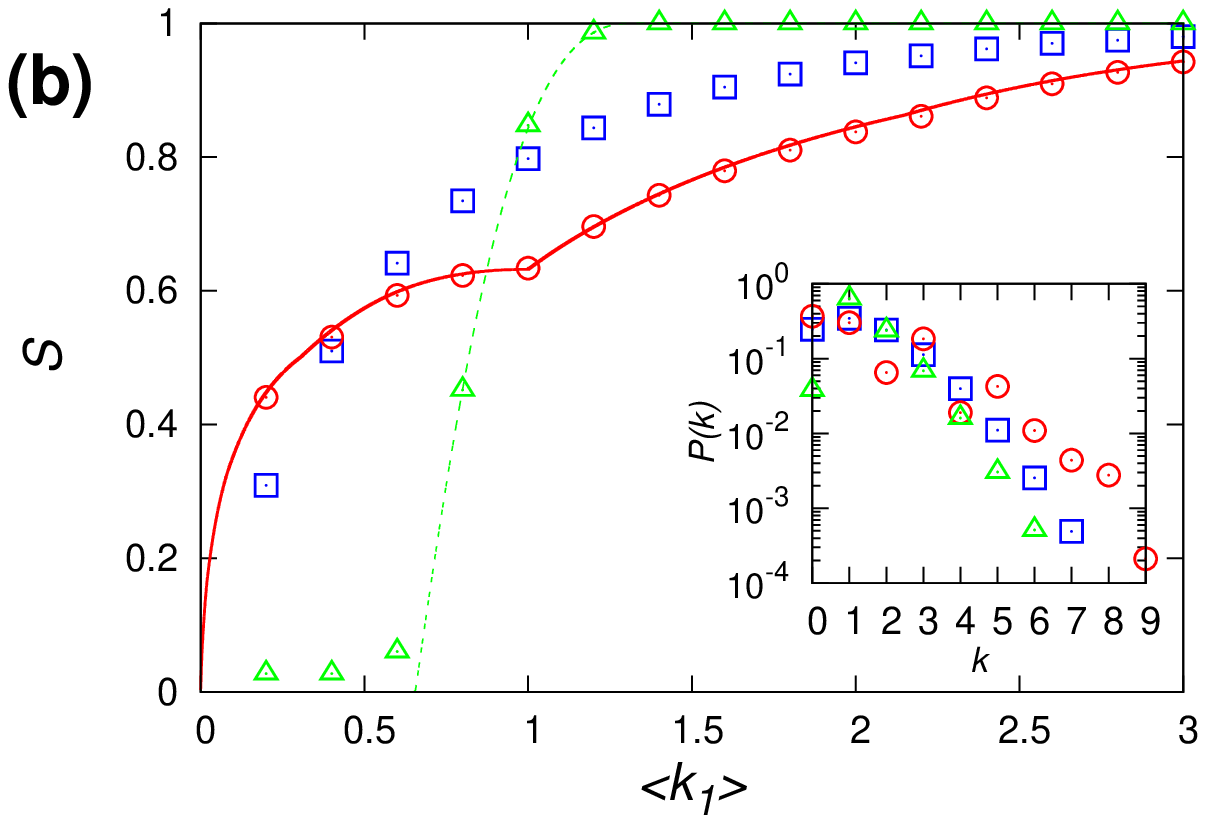}}
\centerline{\epsfxsize=.70\linewidth \epsfbox{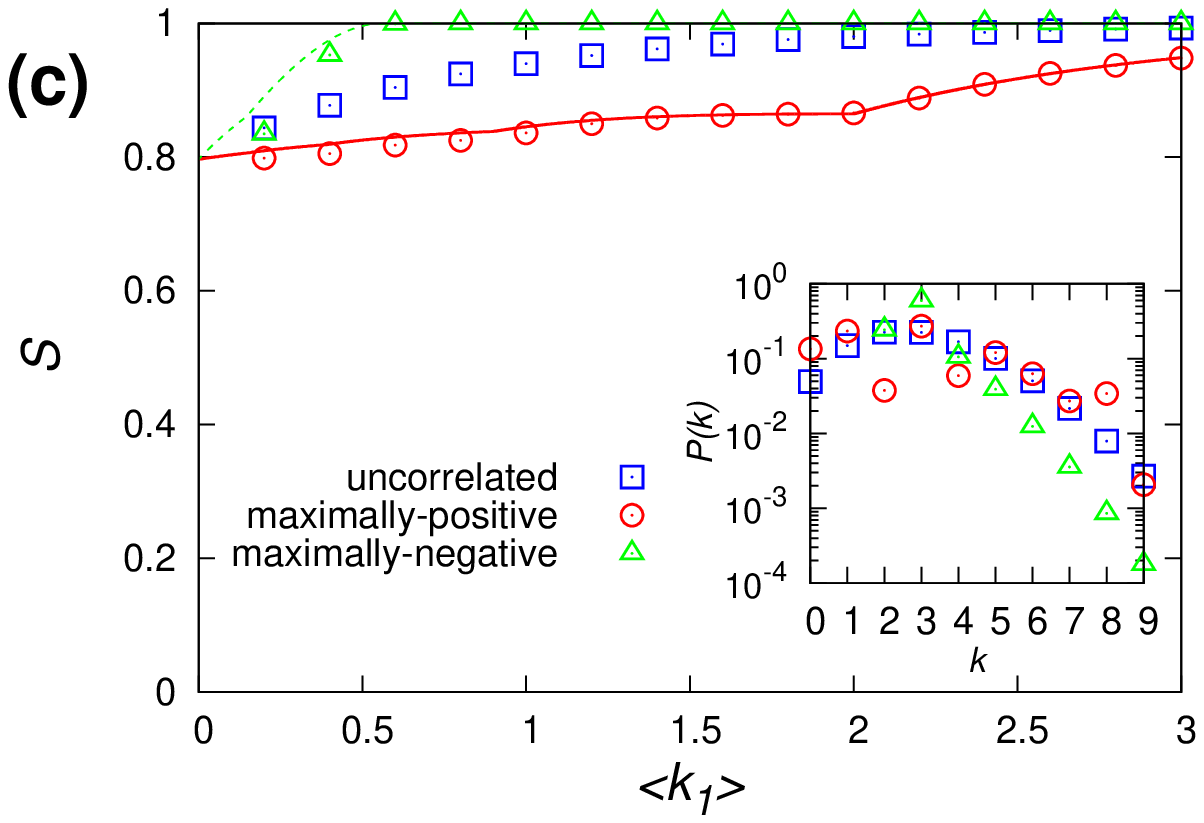}}
\caption{(Color Online) The size of giant component $S$
of the coupled ER networks with two layers, 
with the mean degree of the second layer fixed to be
$\langle k_2\rangle=0.2$ (a), 1.0 (b), and 2.0 (c).
Numerical simulation results for $S$ in uncorrelated ($\Box$), 
maximally-positive ($\circ$), and maximally-negative ($\triangle$) 
cases are presented, along with the mean-field-like calculation results (lines).
(Insets) The degree distribution $P(k)$ of the interlaced networks
in the three cases for $\langle k_1\rangle=1.0$ (a,c) and $0.4$ (b).
}
\end{figure}

Both numerical simulations and mean-field-like calculations show that
in the maximally-positive case, 
the giant component emerges much earlier 
than in the uncorrelated case. 
For example, with $\langle k_2\rangle=0.2$ [Fig.~2(a)], 
which is well below the percolation threshold of a single ER network, 
the giant component emerges at as early as 
$\langle k_1\rangle_c^{(pos)}\approx0.10$ ($\circ$),
much earlier than the uncorrelated case where
it does at $\langle k_1\rangle_c^{(uncorr)}=0.8$ ($\Box$).
This means that we only need less than a third of links 
($\langle k\rangle_c^{(pos)}\sim0.3$ vs.\ $\langle k\rangle_c^{(uncorr)}=1$)
to attain global connectivity in the maximally-positive case with 
$\langle k_2\rangle=0.2$, 
compared with the uncorrelated case.
By contrast, in the maximally-negative case, the giant component
emerges much later. For example, 
with $\langle k_2\rangle=1$ [Fig.~2(b)], exactly the percolation
threshold of a single ER network,
the giant component does not grow to a finite size until 
$\langle k_1\rangle_c^{(neg)}\approx0.65$ ($\triangle$),
that is, $\langle k\rangle_c^{(neg)}\approx1.65$ at $\langle k_2\rangle=1$,
meaning that, surprisingly, additions of about two thirds more links onto
the giant component at the single network percolation point do not
make it grow any larger.
Therefore we need much more links to attain global connectivity
in the maximally-negative case compared with the uncorrelated case.
In this sense, the positive coupling between layers renders
the coupled network more robust under random failures, whereas
the negative coupling decreases the robustness.
These distinct behaviors arise due to the different shapes 
of degree distribution in each case. 
For positive correlation case, the degree distribution
becomes broader than the uncorrelated case, whereas it gets narrower
in the negative correlation case, leading to smaller degree variance 
(Fig.~2, insets).

Although the giant component emerges earlier with positive 
inter-layer coupling, it grows more gradually than the negative 
correlation case, for which it exhibits a more abrupt increase near
the percolation threshold and faster saturation towards $S\to1$ (Fig.~2). 
This suggests an interesting perspective to network robustness;
on one hand, the coupled network with positive inter-layer correlations
is more robust since it can support global connectivity with lower
density of links, whereas on the other hand, the early impact of link removals
from a well-connected network can be more severe, as $S$ is smaller
for the positive correlation case than for the negative correlation case
in the region $S\gtrsim 0.6$. This behavior can also be seen in Fig.~2(c), 
where the giant component size $S$ is plotted against $\langle k_1\rangle$
with $\langle k_2\rangle=2$, well above the percolation threshold.
In this case, the maximally-negative correlation case exhibits
larger $S$ than the maximally-positive case in the whole range of 
$\langle k_1\rangle$ we studied.
These results are somewhat similar to the effect of degree mixing
pattern in the single network \cite{assortative}.

We performed the same calculations for coupled networks 
with two layers of scale-free (SF) networks 
having identical degree distributions of a power-law form,
$\pi^{(l)}(k_l)\sim k_l^{-\gamma}$, with the degree exponent $\gamma=2.3$ and
$3.0$
using the static model \cite{static}, and obtained qualitatively same results.

Finally, we consider real-world examples of coupled network structures.
The first example is the cellular network of a bacterium called
{\it Mycoplasma pneumoniae} \cite{pneumonia}, in which 
nodes are genes or gene products and links are 
the physical bindings to form protein assemblies in the 
protein-protein interaction layer 
and the sharing of metabolites in reactions catalyzed by two enzymes
in the metabolic layer.
The second example is the world trade network \cite{wtn,garlaschelli}, 
in which the nodes are countries and links are trade relations between them,
with two layers corresponding to commodity trades in the primary and 
the secondary industrial sectors, respectively.

\begin{figure}
\centerline{\epsfxsize=.70\linewidth \epsfbox{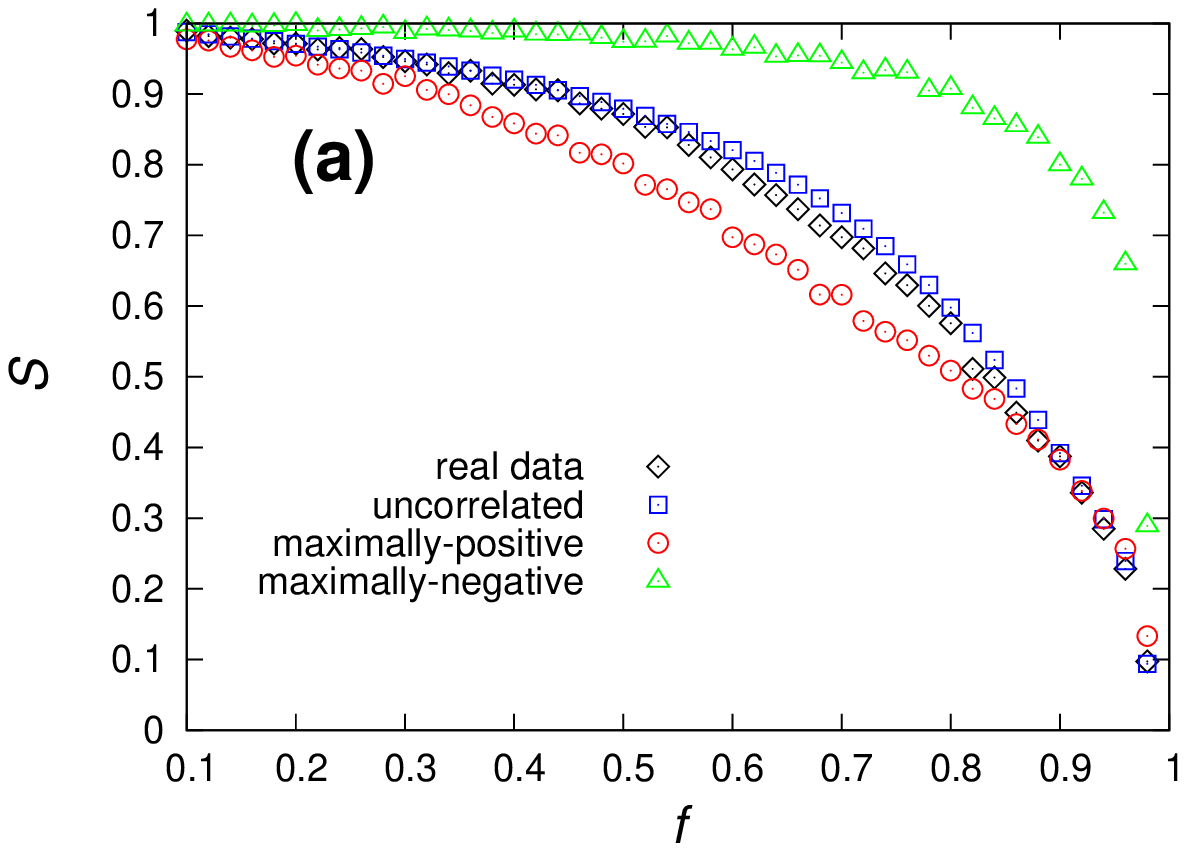}}
\centerline{\epsfxsize=.70\linewidth \epsfbox{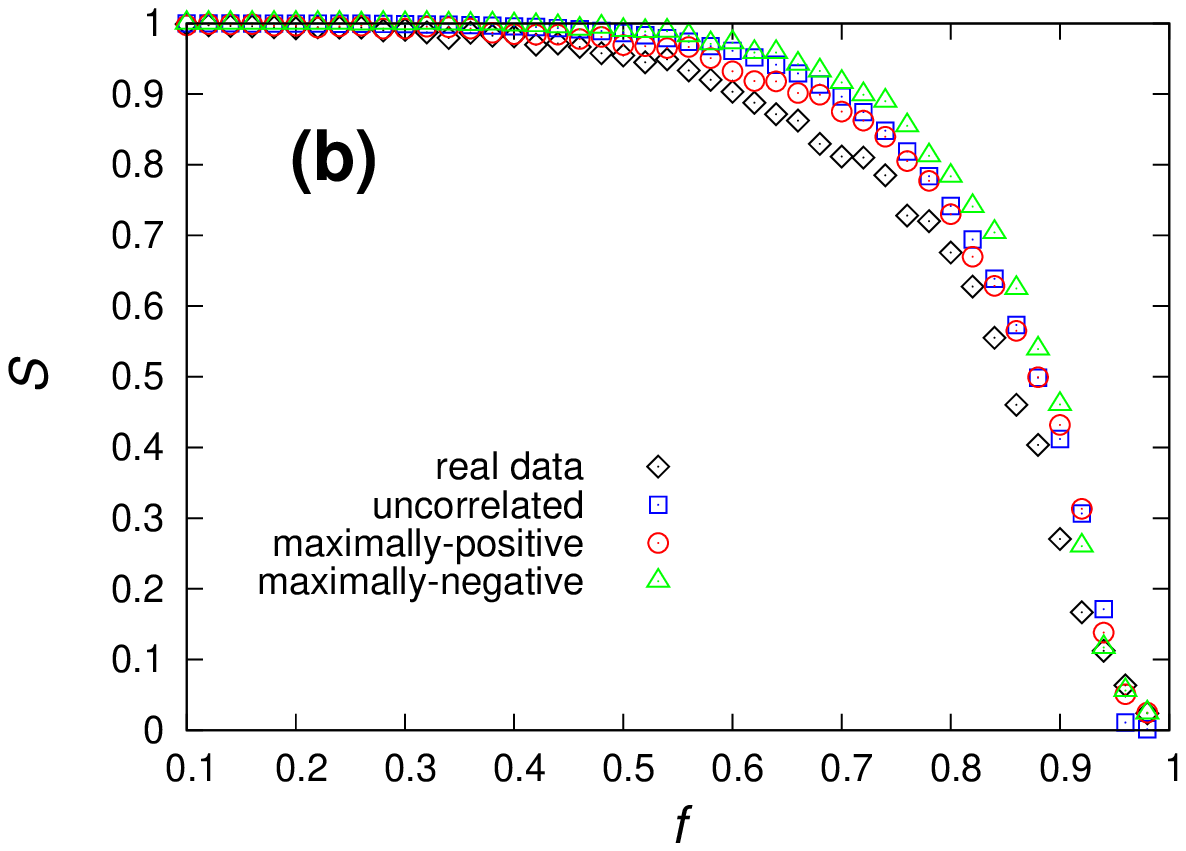}}
\caption{(Color Online) Robustness of real-world coupled networks
under random removal of links. The size of remaining giant component
$S$ after a random removal of fraction $f$ of links ($\diamond$) is plotted
for the cellular network of {\it M. pneumoniae} (a) and the world
trade network (b), along with that for corresponding randomized 
uncorrelated ($\Box$), maximally-positive correlated ($\circ$), and
maximally-negative correlated networks ($\triangle$).
}
\end{figure}

We studied the robustness properties of these coupled networks
under random failures by successively removing links randomly.
In particular we measured the giant component size $S$ as a function
of the fraction of links removed $f$ for the interlaced networks 
(Fig.~3, $\diamond$).
We compared the results with the randomized coupled networks by randomly shuffling
the node identities in the two layers with the intra-layer connectivity intact,
by which we could obtain the uncorrelated coupled networks (Fig.~3, $\Box$).
Furthermore, we constructed the maximally-positive (Fig.~3, $\circ$) 
and maximally-negative (Fig.~3, $\triangle$) correlated coupled networks, 
by ordered-matching and anti-ordered-matching of degree ranks in the two layers 
as in the model networks. As both real-world networks have broad 
intra-layer degree distributions, often of a power-law form with $\gamma<3$, 
both the coupled networks
show extreme resilience under random failures, that is the giant component
sustains ($S>0$) until we remove almost the entire links ($f_c\approx1$).
Distinct characteristics, however, can be observed when we compare them with
corresponding randomized networks. 
The cellular network of {\it M. pneumoniae} behaves
similarly to the uncorrelated randomized networks [Fig.~3(a)], 
whereas the world trade network behaves closely to 
the maximally-positive correlated case [Fig.~3(b)], suggesting a fundamental
difference in the inter-layer coupling structure in these two systems.
Indeed, the inter-layer degree rank correlation is measured to be weak
for the former, $r_s^{cellular}=0.11$, whereas 
it is found to be very strongly positive for the latter, 
$r_s^{trade}=0.86$, consistent
with expectations based on the model network results.

To conclude, we have studied the effect of correlated inter-layer couplings on
the network robustness of coupled network structures, using both 
model networks and real-world networks. It is shown that the positive
inter-layer correlation in degrees renders the network robustness increase
in the sense that the percolation threshold is lowered. At the same time,
the giant component, once formed, grows much more gradually, 
or equivalently, decays much quickly under initial random link removal, 
for positive inter-layer correlations,
which has an important implication for real-world networks with scale-free
topology. We have shown that the correlated coupling between network layers
affects critically the structural properties of coupled network system,
a better representation of most real-world complex systems
than the single, isolated network. Its impact on other dynamic processes
occurring on coupled network systems is to be explored \cite{buldyrev2}.
Finally, higher-order correlation factors within and between network layers,
such as link overlaps, mixing, clustering, and community patterns, 
could also induce other nontrivial impact on coupled network structure and
dynamics, opening a wide avenue for future studies.

\begin{acknowledgments}
We thank J. Choi and K.-M. Lee for their help 
with real-world network data.
This work was supported by Mid-career Researcher Program
through NRF grant funded by the MEST (No.~2009-0080801).
While finalizing the manuscript, we learned of an independent
study by Parshani {\it et al.} \cite{inter-similarity} 
which dealt with a similar problem on interdependent networks
and obtained results partly overlapping with ours.
\end{acknowledgments}


\begin{thebibliography}{99}
\bibitem{foot} $r_s$ becomes strictly $1$ or $-1$ for maximally-correlated cases
when there are no ties in the ranks. When there are ties, the average
rank is used to compute $\Delta_i$.
\end{thebibliography}
\end{document}